\documentstyle[12pt]{article}

\catcode`\@=11
\long\def\@makefntext#1{ 
\protect\noindent \hbox to 3.2pt {\hskip-.9pt
$^{{\ninerm\@thefnmark}}$\hfil}#1\hfill} 

\def\thefootnote{\fnsymbol{footnote}}
 \def\@makefnmark{\hbox to 0pt{$^{\@thefnmark}$\hss}}  

\def\ps@myheadings{\let\@mkboth\@gobbletwo
\def\@oddhead{\hbox{} 
\rightmark\hfil\ninerm\thepage}
\def\@oddfoot{}\def\@evenhead{\ninerm\thepage\hfil 
\leftmark\hbox{}}\def\@evenfoot{}
\def\sectionmark##1{}\def\subsectionmark##1{}}

\newcommand{\be}{\begin{equation}} \newcommand{\ee}{\end{equation}}
\newcommand{\ba}{\begin{eqnarray}} \newcommand{\ea}{\end{eqnarray}}

\begin{document}

\newcommand{\symbolfootnote}{\renewcommand{\thefootnote}
	{\fnsymbol{footnote}}}
\renewcommand{\thefootnote}{\fnsymbol{footnote}}
\newcommand{\alphfootnote}
	{\setcounter{footnote}{0}
	 \renewcommand{\thefootnote}{\sevenrm\alph{footnote}}}

\newcounter{sectionc}\newcounter{subsectionc}\newcounter{subsubsectionc}
\renewcommand{\section}[1] {\vspace{0.6cm}\addtocounter{sectionc}{1}
\setcounter{subsectionc}{0}\setcounter{subsubsectionc}{0}\noindent
	{\bf\thesectionc. #1}\par\vspace{0.4cm}}
\renewcommand{\subsection}[1] {\vspace{0.6cm}\addtocounter{subsectionc}{1}
	\setcounter{subsubsectionc}{0}\noindent
	{\it\thesectionc.\thesubsectionc. #1}\par\vspace{0.4cm}}
\renewcommand{\subsubsection}[1]
{\vspace{0.6cm}\addtocounter{subsubsectionc}{1}
	\noindent {\rm\thesectionc.\thesubsectionc.\thesubsubsectionc.
	#1}\par\vspace{0.4cm}}
\newcommand{\nonumsection}[1] {\vspace{0.6cm}\noindent{\bf #1}
	\par\vspace{0.4cm}}

\newcounter{appendixc}
\newcounter{subappendixc}[appendixc]
\newcounter{subsubappendixc}[subappendixc]
\renewcommand{\thesubappendixc}{\Alph{appendixc}.\arabic{subappendixc}}
\renewcommand{\thesubsubappendixc}
	{\Alph{appendixc}.\arabic{subappendixc}.\arabic{subsubappendixc}}

\renewcommand{\appendix}[1] {\vspace{0.6cm}
        \refstepcounter{appendixc}
        \setcounter{figure}{0}
        \setcounter{table}{0}
        \setcounter{equation}{0}
        \renewcommand{\thefigure}{\Alph{appendixc}.\arabic{figure}}
        \renewcommand{\thetable}{\Alph{appendixc}.\arabic{table}}
        \renewcommand{\theappendixc}{\Alph{appendixc}}
        \renewcommand{\theequation}{\Alph{appendixc}.\arabic{equation}}
        \noindent{\bf Appendix \theappendixc #1}\par\vspace{0.4cm}}
\newcommand{\subappendix}[1] {\vspace{0.6cm}
        \refstepcounter{subappendixc}
        \noindent{\bf Appendix \thesubappendixc. #1}\par\vspace{0.4cm}}
\newcommand{\subsubappendix}[1] {\vspace{0.6cm}
        \refstepcounter{subsubappendixc}
        \noindent{\it Appendix \thesubsubappendixc. #1}
	\par\vspace{0.4cm}}

\def\abstracts#1{{
	\centering{\begin{minipage}{30pc}\tenrm\baselineskip=12pt\noindent
	\centerline{\tenrm ABSTRACT}\vspace{0.3cm}
	\parindent=0pt #1
	\end{minipage} }\par}}

\newcommand{\bibit}{\it}
\newcommand{\bibbf}{\bf}
\renewenvironment{thebibliography}[1]
	{\begin{list}{\arabic{enumi}.}
	{\usecounter{enumi}\setlength{\parsep}{0pt}
\setlength{\leftmargin 1.25cm}{\rightmargin 0pt}
	 \setlength{\itemsep}{0pt} \settowidth
	{\labelwidth}{#1.}\sloppy}}{\end{list}}

\topsep=0in\parsep=0in\itemsep=0in
\parindent=1.5pc

\newcounter{itemlistc}
\newcounter{romanlistc}
\newcounter{alphlistc}
\newcounter{arabiclistc}
\newenvironment{itemlist}
    	{\setcounter{itemlistc}{0}
	 \begin{list}{$\bullet$}
	{\usecounter{itemlistc}
	 \setlength{\parsep}{0pt}
	 \setlength{\itemsep}{0pt}}}{\end{list}}

\newenvironment{romanlist}
	{\setcounter{romanlistc}{0}
	 \begin{list}{$($\roman{romanlistc}$)$}
	{\usecounter{romanlistc}
	 \setlength{\parsep}{0pt}
	 \setlength{\itemsep}{0pt}}}{\end{list}}

\newenvironment{alphlist}
	{\setcounter{alphlistc}{0}
	 \begin{list}{$($\alph{alphlistc}$)$}
	{\usecounter{alphlistc}
	 \setlength{\parsep}{0pt}
	 \setlength{\itemsep}{0pt}}}{\end{list}}

\newenvironment{arabiclist}
	{\setcounter{arabiclistc}{0}
	 \begin{list}{\arabic{arabiclistc}}
	{\usecounter{arabiclistc}
	 \setlength{\parsep}{0pt}
	 \setlength{\itemsep}{0pt}}}{\end{list}}

\newcommand{\fcaption}[1]{
        \refstepcounter{figure}
        \setbox\@tempboxa = \hbox{\tenrm Fig.~\thefigure. #1}
        \ifdim \wd\@tempboxa > 6in
           {\begin{center}
        \parbox{6in}{\tenrm\baselineskip=12pt Fig.~\thefigure. #1 }
            \end{center}}
        \else
             {\begin{center}
             {\tenrm Fig.~\thefigure. #1}
              \end{center}}
        \fi}

\newcommand{\tcaption}[1]{
        \refstepcounter{table}
        \setbox\@tempboxa = \hbox{\tenrm Table~\thetable. #1}
        \ifdim \wd\@tempboxa > 6in
           {\begin{center}
        \parbox{6in}{\tenrm\baselineskip=12pt Table~\thetable. #1 }
            \end{center}}
        \else
             {\begin{center}
             {\tenrm Table~\thetable. #1}
              \end{center}}
        \fi}

\def\@citex[#1]#2{\if@filesw\immediate\write\@auxout
	{\string\citation{#2}}\fi
\def\@citea{}\@cite{\@for\@citeb:=#2\do
	{\@citea\def\@citea{,}\@ifundefined
	{b@\@citeb}{{\bf ?}\@warning
	{Citation `\@citeb' on page \thepage \space undefined}}
	{\csname b@\@citeb\endcsname}}}{#1}}

\newif\if@cghi
\def\cite{\@cghitrue\@ifnextchar [{\@tempswatrue
	\@citex}{\@tempswafalse\@citex[]}}
\def\citelow{\@cghifalse\@ifnextchar [{\@tempswatrue
	\@citex}{\@tempswafalse\@citex[]}}
\def\@cite#1#2{{$\null^{#1}$\if@tempswa\typeout
	{IJCGA warning: optional citation argument
	ignored: `#2'} \fi}}
\newcommand{\citeup}{\cite}

\def\fnm#1{$^{\mbox{\scriptsize #1}}$}
\def\fnt#1#2{\footnotetext{\kern-.3em
	{$^{\mbox{\sevenrm #1}}$}{#2}}}

\font\twelvebf=cmbx10 scaled\magstep 1
\font\twelverm=cmr10 scaled\magstep 1
\font\twelveit=cmti10 scaled\magstep 1
\font\elevenbfit=cmbxti10 scaled\magstephalf
\font\elevenbf=cmbx10 scaled\magstephalf
\font\elevenrm=cmr10 scaled\magstephalf
\font\elevenit=cmti10 scaled\magstephalf
\font\bfit=cmbxti10
\font\tenbf=cmbx10
\font\tenrm=cmr10
\font\tenit=cmti10
\font\ninebf=cmbx9
\font\ninerm=cmr9
\font\nineit=cmti9
\font\eightbf=cmbx8
\font\eightrm=cmr8
\font\eightit=cmti8

\centerline{\tenbf INCLUSIVE AND SEMI-INCLUSIVE DEEP INELASTIC SCATTERING}
\baselineskip=22pt
\centerline{\tenbf AT CEBAF AT HIGHER ENERGIES}
\baselineskip=16pt
\vspace{0.8cm}
\centerline{\tenrm B. FROIS}
\baselineskip=13pt
\centerline{\tenit Service de Physique Nucl\'eaire, Centre d'Etudes de Saclay}
\baselineskip=12pt
\centerline{\tenit F-91191 Gif sur Yvette CEDEX, France}
\vspace{0.3cm}
\centerline{\tenrm and}
\vspace{0.3cm}
\centerline{\tenrm P.J. MULDERS}
\baselineskip=13pt
\centerline{\tenit National Institute for Nuclear Physics and High Energy
Physics}
\baselineskip=12pt
\centerline{\tenit (NIKHEF-K),
P.O. Box 41882, NL-1009 DB Amsterdam, the Netherlands}
\vspace{0.9cm}
\abstracts{
We summarize the discussion on the possibilities of doing inclusive and
semi-inclusive deep inelastic scattering experiments at CEBAF with
beam energy of the order of 10 GeV.
}

\vspace{0.9cm}
\vfil
\rm\baselineskip=14pt

This summary is based on talks by A. Mueller and
A. Schaefer and contributions of N. Bianchi, X. Ji, J.-M. Laget, P.
Markowitz, W. Melnitchouk, Z.-E. Meziani, J. Milana, P. Mulders, S.
Simula, P. Souder and M. Strikman. These contributions can be found
elsewhere in this report.

Electron scattering  from a  composite target  is due  to an electroweak
interaction,  well-described  by  the exchange of a
virtual boson, either a photon or $Z^0$.  The exchanged particle carries
a momentum $q$ and probes the distribution of electroweak charges in the
target which are carried  by the (elementary) constituents,  the quarks.
Assuming  that  the  momentum  of  the  target  is  $P$,  the
invariant variables are the momentum transfer $Q^2$ and the energy
transfer $\nu$
\begin{eqnarray}
& & Q^2 = - q^2 \ \stackrel{TRF}{=} \ 4 E_e E_e^\prime
\,\sin^2(\theta_e/2),\\
& & \nu = \frac{P\cdot q}{M} \ \stackrel{TRF}{=} \
E_e - E_e^\prime.
\end{eqnarray}
The momentum transfer is a spacelike  vector.
In the laboratory frame or target-rest-frame (TRF) these are
simply determined
by measuring the momentum of the incident and scattered electrons.

Deep inelastic scattering assumes  that the electron resolves  the quark
structure of the  nucleon.  This  requires a sufficiently  high momentum
transfer to be  able to understand  the reaction mechanism  in terms of
perturbative  QCD  process.    A  momentum  transfer  $Q  >  1$ to 2
GeV is considered as the typical lower boundary for deep inelastic
scattering. Transforming this into a spatial resolution, $\lambda
\approx 1/Q$ gives $\lambda$ = 0.1 - 0.2 fm.  The deep inelastic
scattering corresponds to a an incoherent sum over  scattering
off the individual quarks  weighted by the  squared charges.   Besides
the momentum  transfer, one requires that the invariant mass $W$ of  the
hadronic system is above the  region of discrete baryon resonances, $W
\ge$ 2 GeV.  The invariant mass $W$ is given by

\begin{equation}
W^2 = (P+q)^2 = M^2 + 2\,M\nu - Q^2 = M^2 + \frac{1-x}{x}\,Q^2,
\end{equation}

where $x$ is the  Bjorken scaling variable $x$  = $Q^2/2P\cdot q$.   The
variable  $x$  can  be  interpreted  as  the  fraction  of the lightcone
momentum of the struck quark compared to that of the nucleon,

\begin{equation}
x = \frac{p^+}{P^+}
\end{equation}

Note that $p$ is the quark momentum and $p^+$ = $(p^0 +  p^3)/\sqrt{2}$.
The two  constraints ($Q  > 1  $~GeV and  $W \ge$  2 GeV ) restricts the
experimental explorations  at CEBAF  to $1  \le Q^2  \le 10$ GeV$^2$ and
$0.1 \le x \le  1$ for deep inelastic  scattering on a nucleon.
Fig.~\ref{KIN}
\begin{figure}[t]
\vspace{8 true cm}
\caption
{\protect\small \protect\sl
The kinematic $Q^2 - \nu$ region accessible with a beam
energy of 8 - 12 GeV (left of indicated lines).
Shown are lines of constant $x$ and constant $W$.
}
\label{KIN}
\end{figure}
shows the the specific domain  accessible at CEBAF.
Shown are furthermore lines of constant $x$ for $x$ = $10^{-2}$,
$10^{-1}$, $1$ (elastic scattering off a nucleon), $200$ (elastic scattering
off a heavy nucleus with $A$ = 200), constant $W$ = $M_\Delta$ (dashed,
labeled $\Delta$), 2 GeV (dashed, labeled $N^\ast$) and $W$ corresponding to
the threshold for $J/\psi$ production (dashed, labeled $c$). The {\em region
of deep inelastic scattering} is bounded by $W$ = 2 GeV and the horizontal
dot-dashed line corresponding with $Q^2$ = 1 GeV$^2$. The region between the
(dashed) lines for $W$ = $M_\Delta$ and $W$ = 2 GeV is the {\em resonance
region}. The right edge of the figure
corresponds to $\nu$ = 200 GeV, which is about the maximum energy transfer
in the CERN muon experiments.
The  measurement of
the  valence  quark  structure  of  hadrons and the correlations between
quarks  and  gluons  by  deep  inelastic  studies  have been extensively
discussed in the review of Sloan,  Smadja and Voss ~\cite{Sloan}, in
the Pegasys proposal~\cite{PEGASYS} and the ELFE proposal~\cite{ELFE}.

\section{The nucleon spin structure}

In the  study  of  valence  quarks,  which  carry a sizable fraction of
the momentum of the nucleon, an illustrative example of the
possibilities  at CEBAF  is  the  determination  of  the  precise
shape  of the polarized structure functions $g_1^p$  and $g_1^n$ at
large $x$-values.   Present available data on  the neutron spin
structure function at  high Bjorken variable  $x$  stop  at  a  value
of  0.5.    The precise shape and the threshold behavior for  $x
\rightarrow 1$  does not affect  the sum rule but it {\em is} important
for the comparison with model calculations  of the quark distributions.

In order to test the prediction $A_1\rightarrow 1$ as $ x\rightarrow  1$
precision data at  $x>0.5$ are necessary.   Furthermore, the  slope with
which the asymmetry  reaches unity becomes  sensitive to the  details of
the nucleon  substructure.   Severals constituent  quark models although
constrained by  the limit  at $x=1$,  produce an  asymmetry $A_1$  which
behaves differently in the large $x$ region.

{}From the  experimental point  of view,  measurements of  $A_1$ at high
energy are  not feasible,  because high  $x$ implies  high $E^{\prime }$
(usually more that 10 GeV) the resolution of the  high energy
spectrometers does not  allow a precise  measurement of the  slope since
data are averaged over a wide  $x$ range.  The low count  rate conspires
to make the measurement extremely difficult.

Therefore, it has been proposed to use the combination of unique
possibilities of a high  resolution  8   GeV  incident  electron   beam
and  the   Hall  A spectrometers at large scattering angles.   This
implies a large  recoil energy and therefore small $E'$, typically below
4 GeV.  The  high momentum resolution of  Hall A spectrometers
allows to step  in the $x$ region between 0.2 and 0.7 finely keeping
$W^2$ greater then 4 GeV$^2  $ to insure that the scattering  process is
in the deep  inelastic region. The depolarization  factor for  a given
$x$ is  closer to  one at large angle putting CEBAF at advantage  for the
measurement using large  angle since  $A_1=A_{1\parallel}/D$.    Two
identical spectrometers with 10\% momentum acceptance allow to double
the solid angle from 7.5 to 15  msr. When combined with  a 40 cm  long
high pressure  polarized $^3$He target and a 15 $\mu $A electron beam
this setup offers a superb luminosity  of $10^{36}/$cm$^2/s$.

The proposal discussed by Meziani at this workshop showed that  in 1000
hours of beam time one can achieve a precise measurement of the  neutron
spin structure function at high $x$.

\section{Parity violation}

It was proposed that a  parity  violation  experiment  in  the deep
inelastic  region  would  be  of  special interest for $x=0.5$, $y=0.5$.
Such  a  measurement  would  allow  a  precise  test of the axial-hadron
vector-electron electroweak  coupling.   The particular  point
above will  be less sensitive to  structure functions, while
measurements for lower  x and  y  values  are  expected  to  provide
information  about structure functions.

\section{Nonleading structure functions}

The ratio of longitudinal and transverse cross sections  $R(x,Q^2)$ (at
relatively small  $Q^2$) is  nonzero as  a result  from a combination of
perturbative  QCD  corrections,  which   are  sensitive  to  the   gluon
distribution, higher twist corrections such as target mass  corrections,
and finally  nonperturbative effects.   Besides  the study  of $R$ for a
nucleon target, the study of $R_A$ in a nucleus is of interest.

Another topic which we mention is the study of higher twist structure
functions. These are of interest because  they are a manifestation of
correlations between quarks or between quarks and gluons.  These
correlations  vanish as powers of $1/Q$ and are therefore more prominent
at relatively  small values of $Q^2$  (although one must  be in the
deep inelastic region!). In  inclusive  unpolarized  electron
scattering  the first higher twist contribution  is  ${\cal
O}(1/Q^2)$.    At ${\cal O}(1/Q)$ higher twist contribution  can  be
measured  by  using  polarized  beams  and/or the detection of produced
hadrons.  As compared to the leading  contribution the  higher  twist
contributions  turn  out  to  be  stronger at larger $x$-values, as
shown in NMC experiments at CERN.

\section{Semi-inclusive experiments}

A region, where deep inelastic experiments are less restricted by energy
considerations is the region of $x \ge 1$, accessible in scattering  off
nuclei.  Here the beam luminosity  is much more important, as the  cross
sections are small.  After all, one is looking at one out of six valence
quarks which carries a large fraction (more than 1/2) of the momentum of
two nucleons, or one is looking at a quark belonging to a  high-momentum
nucleon in the nucleus.  The  distinction between these two can be  made
as in  the second  process one  can try  to detect another high-momentum
nucleon or nuclear rest-system moving in the backward direction.   These
type  of  experiments   (tagged  structure  functions)   requiring  high
luminosity and special detector set-ups are well suited for an  upgraded
CEBAF.

The case of tagged structure functions is an example of a semi-inclusive
process in  which one  (or more)  particles are  detected in coincidence
with  the  scattered  electron.     For  semi-inclusive  processes   one
distinguishes  different  production  mechanisms  such  as  {\em  target
fragmentation} and  {\em current  fragmentation}.   The latter  case, in
which one is interested mainly in the particles produced in the  forward
direction, can at sufficiently high  energies be described as a  product
of  quark  distribution  functions  $f_{H\rightarrow  q}(x)$  and  quark
fragmentation  functions  $D_{q\rightarrow  h}(z)$.    Here  $z = P\cdot
P_h/P\cdot q$, which in the target rest-frame is $z = E_h/\nu$, i.e. the
fraction of the energy  of the photon or  the struck quark taken  by the
produced hadron.   The fragmentation can  be used to  tag specific quark
flavors or their spins, e.g. an $s$-quark will favor production of $K^-$
(or $\overline K^0$) while an $\overline s$-quark will favor  production
of  $K^+$  (or  $K^0$),  leading  to  asymmetries  in hadroproduction of
specific particles  (that must  then be  identified).   In order for the
factorization to be  valid and have  a sufficiently clear  separation of
the target and  current fragmentation region,  an energy of  $10$ GeV is
too low\cite{Sloan}.

\section{Conclusions}

The region, which for CEBAF is most important is the transition  between
the  (inclusive)  deep  inelastic  scattering  region  and the region of
exclusive processes, such as the excitation of baryon resonances leading
to a specific final state.  In exclusive processes the momentum transfer
in the $t$-channel  becomes a relevant  variable that determines  if one
can use perturbative QCD methods to describe the process.  In this  case
the object of study are not  the quark probabilities in the target,  but
one is sensitive to the  (lightcone) quark wave functions, which  depend
on  the  lightcone  momentum  fractions  $x_i$  =  $p^+_i/P^+$  and  the
perpendicular  quark  momenta.    Focussing  on  elastic  processes   or
exclusive or semi-inclusive production of vector mesons ($\rho$,  $\phi$
or  $J/\psi$  one  can  investigate  different  components  of  the wave
functions or specific  reaction mechanisms.

An upgrade of  energy up to  about $10$ GeV  gives a relatively  limited
access to  the region  of inclusive  deep inelastic  scattering and even
less  to  the  region  where  semi-inclusive  deep  inelastic  processes
factorize.   However, this  workshop has  shown that  such an upgrade of
CEBAF  beam  energy  would  allow  to  address  a  number of fundamental
questions  related  to  the  valence  quark  structure  of hadrons.  The
answers could clarify  some puzzles related  to the approach  to scaling
even at modest $Q^2$-values.  For this a broad coverage of the resonance
region reaching into the deep inelastic scattering region is  important.
It is here where the challenge is to connect effective hadronic theories
or  successful  quark  models  to  the  underlying  theory  of   quantum
chromodynamics.

This work is in part (P.J.M) supported by the foundation for Fundamental
Research of Matter  (FOM) and the  National Organization for  Scientific
Research (NWO).

\end{document}